\newcommand{\bee}{\begin{equation}}
\newcommand{\ene}{\end{equation}}

\documentclass[pre,twocolumn,showpacs,amsmath,amssymb]{revtex4}
\usepackage[]{graphicx}
\usepackage{color}

\begin{document}


\title{Quasi-two-dimensional complex plasma containing spherical particles and their binary agglomerates}

\author{M. Chaudhuri$^{1*}\footnotetext{* corresponding author: manischaudhuri@g.harvard.edu}$, I. Semenov$^2$, V. Nosenko$^2$, H. M. Thomas$^2$}
\affiliation{$^1$School of Engineering and Applied Sciences (SEAS), Harvard University, USA}
\affiliation{$^2$Forschungsgruppe Komplexe Plasmen,
Deutsches Zentrum f\"{u}r Luft- und Raumfahrt, D-82234, We{\ss}ling, Germany}

\begin{abstract}
A new type of quasi-two-dimensional complex plasma system was observed which consisted of monodisperse microspheres and their binary agglomerations (dimers).~The particles and their dimers levitated in a plasma sheath at slightly different heights and formed two distinct sublayers.~The system did not crystallize and may be characterized as disordered solid.~The dimers were identified based on their characteristic appearance in defocused images, i.e., rotating interference fringe patterns. The in-plane and inter-plane particle separations exhibit nonmonotonic dependence on the discharge pressure which agrees well with theoretical predictions.
\end{abstract}


\maketitle

Complex plasma is considered as a unique model system in soft matter physics where solid charged particles (dust) form a suspension in the background of weakly ionized plasma~\cite{Morfill_RMP,Morfill:lowen,Morfill:pop,Chaudhuri_SM}. Typically, the dust particles are large enough ($\sim \mu$m) to be visualized individually which allow experimental investigations with high temporal and spatial resolution in terms of the appropriate plasma frequency and particle separation. The dynamics of each particle can be tracked precisely after analyzing the image sequences taken by video camera. In the strong coupling regime, complex plasma forms ``crystals'' which can be arranged in 2D and 3D structures~\cite{Thomas-Nature-1996,Fortov2005PR,Chu_PRL,Thomas_PRL}. In particular, the 2D crystal has been used extensively to investigate various generic processes associated with strong coupling phenomena such as crystallization~\cite{Hartmann_PhysRevLett.105.115004}, melting~\cite{Quinn_PhysRevE.64.051404,Nosenko_PhysRevLett.93.155004,Nosenko_PhysRevLett.103.015001,Feng_PhysRevLett.104.165003}, dislocation dynamics~\cite{Nosenko_PhysRevLett.99.025002}, recrystallization~\cite{Knapek_PhysRevLett.98.015004}, Mach cones~\cite{Samsanov_PhysRevE.61.5557}, dynamical heterogeneity~\cite{Juan_PhysRevE.64.016402}, transport phenomena such as heat conductivity~\cite{Nunomura_PhysRevLett.95.025003}, viscosity~\cite{Nosenko_PhysRevLett.93.155004}, superdiffusion~\cite{Liu_PhysRevLett.100.055003}, etc. Quasi-2D systems consisting of two different types of monodisperse particles have also been explored~\cite{ivlev_PhysRevX,Hyde_PhysRevE}. Additionally, kinetics of phase transition such as crystallization front propagation~\cite{Zuzic:Naturephysics}, fluid-solid transition under microgravity~\cite{Khrapak:PhysRevLett.106.205001,Khrapak:PhysRevE.85.066407}, lane formation~\cite{Du:1367-2630-14-7-073058}, etc have been investigated in 3D plasma crystals.

\begin{figure}[h]
\includegraphics[width=\linewidth]{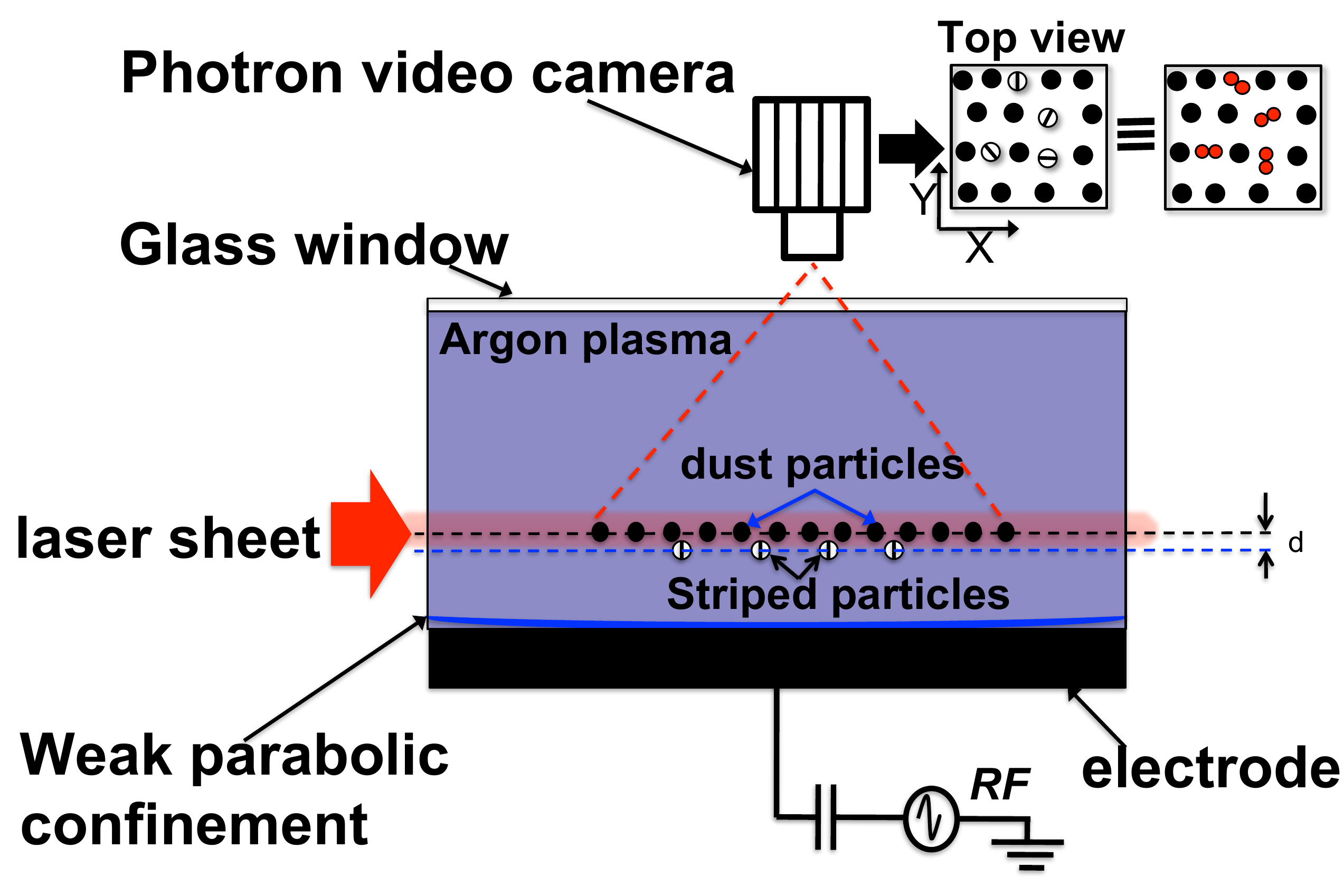}
\caption{Sketch of the experimental setup after partial purification. The microparticles are confined in the weak parabolic confinement potential above the rf electrode and are illuminated with a horizontal laser sheet (width, D $\sim$ 180 $\mu$m). The dimers can be identified instantly by looking at the defocused top-view images (striped particles) but in the focused image they look brighter compared to spherical particles. However, actual orientation of the dimers can not be resolved with present diagnostic. They levitate just below the monolayer formed by the spherical particles (black particles) without forming vertical pairs. Both types of particles can be observed by top view camera. The vertical separation d can be measured by adjusting the maximum laser intensity on two different planes of spherical particles and dimers separately as d $>$ D.}
\label{GEC} 
\end{figure} 
The goal of this paper is to explore the structural properties of a quasi-2D suspension (a mixture of spherical particles and their binary agglomerates), which can be created at the centre using controlled systematic partial purification method. Defocused imaging technique has been used as a simple but powerful diagnostic to identify binary agglomerates in complex plasmas~\cite{APL_Manis}. Unlike spherical particles, the defocused images of  binary agglomerates (two spherical particles together or dimers) contain stationary/spinning interference fringe patterns. This technique has been exploited to create quasi-2D suspension as a binary mixture of normal spherical particles and their binary agglomerates using only one type of particles (of specific size) in a controlled process. In quasi-2D systems, the vertical separation between particles (d) is much less than their in-plane interparticle separation $(\Delta)$ and the top view image should contain all the particles (i.e., there should not be vertical pairing). The static structural properties of both regions have been compared. It is found that the presence of binary agglomerates makes the quasi-2D suspension deformed disordered solid. The deformation (ratio of horizontal and vertical interparticle separations) has been measured over a wide range of neutral gas pressure.

\begin{figure}
\includegraphics[width=\linewidth]{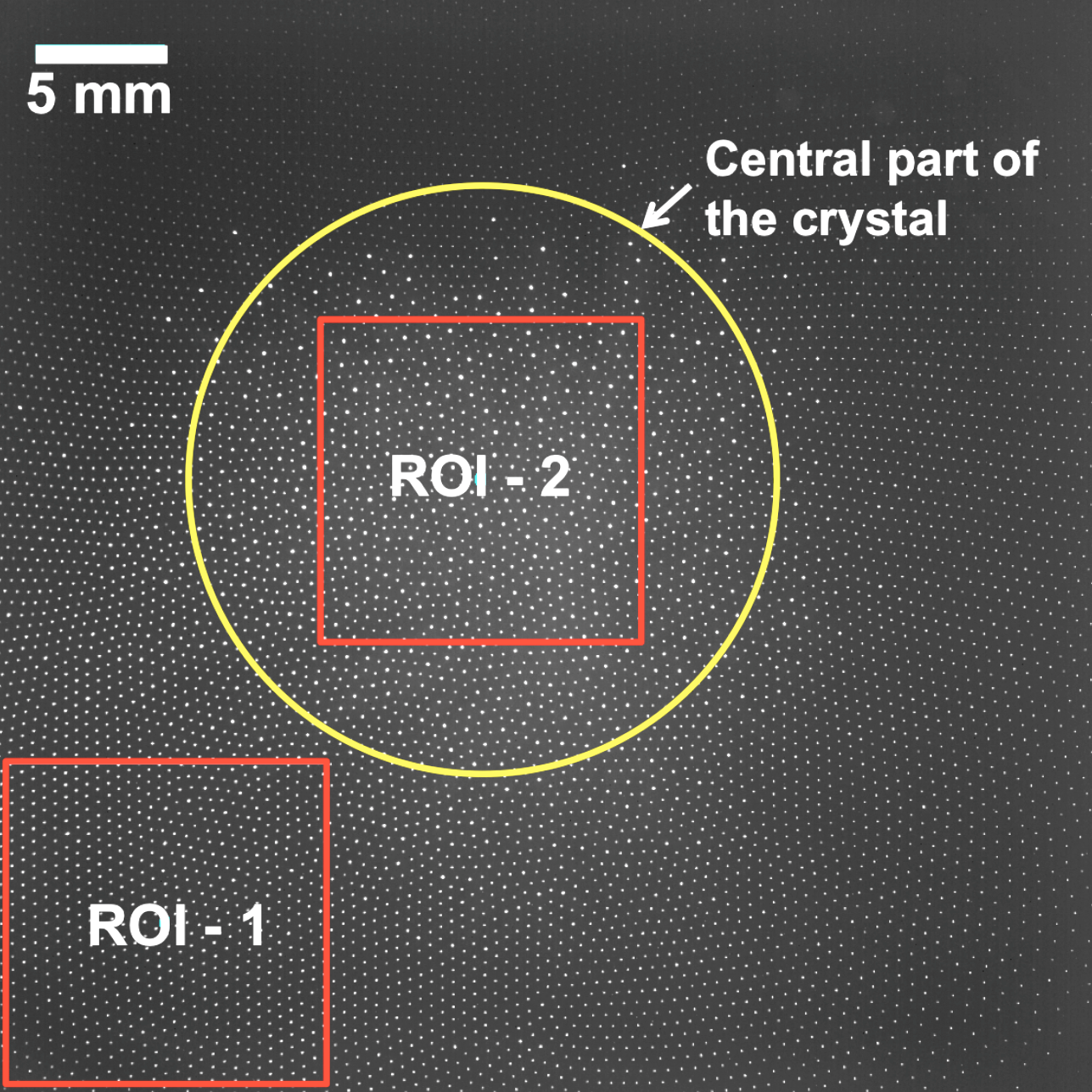}
\caption{(a) Top-view image of the suspension after controlled partial purification which exhibits the presence of both quasi-2D suspension (binary mixture of spherical particles and their binary agglomerates) in the centre and 2D suspension at the periphery. The agglomerates concentrate mainly in the central part of the crystal because of their higher mass and the depth of the confining potential. Two regions of interest (ROI) have been considered of same dimensions (300 pixels x 300 pixels) for analysis: ROI-1 (periphery) and ROI-2 (center).}
\label{image_patterns} 
\end{figure}

\begin{figure}
\includegraphics[width=\linewidth]{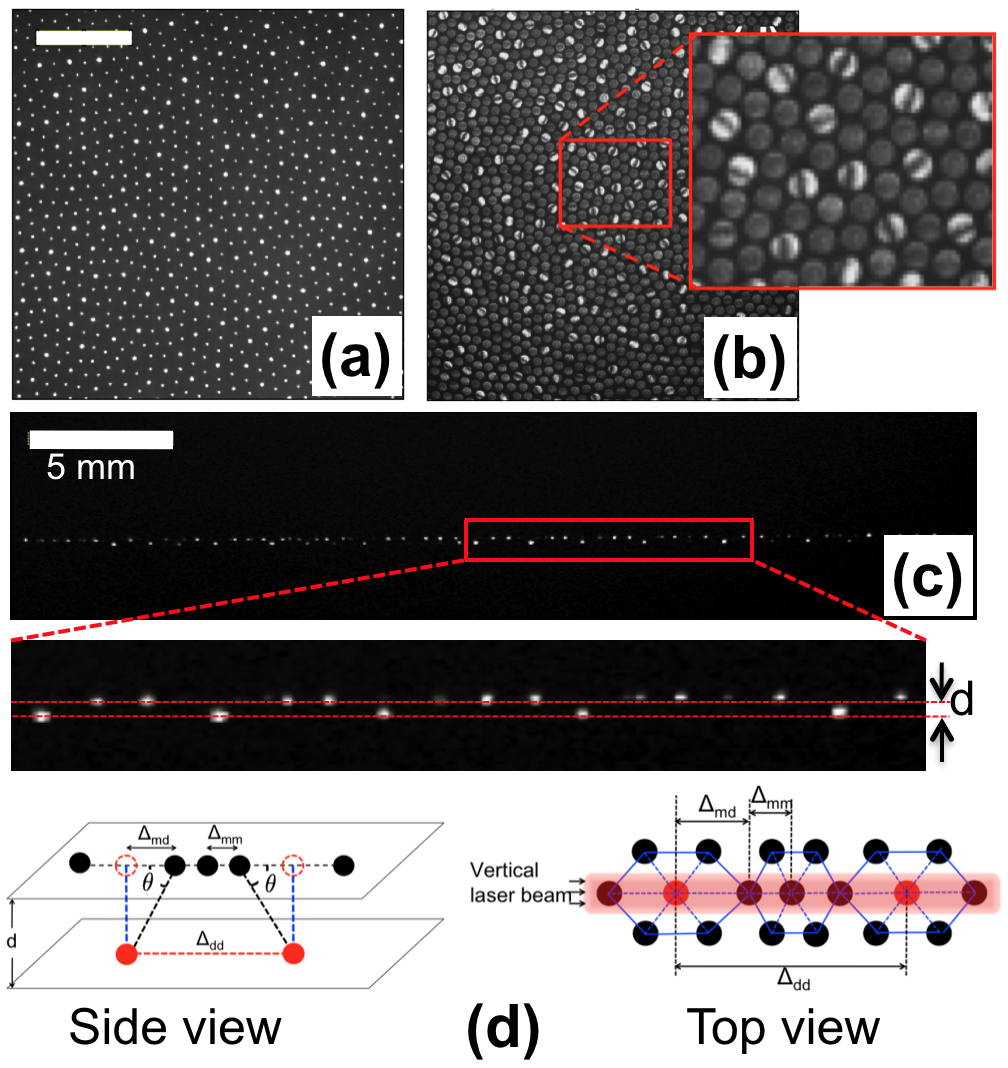}
\caption{Identification test for the binary agglomerates. (a) Focused and (b) defocused image of quasi-2D suspension after the controlled partial purification method. The maximum intensity of the laser beam was placed on the dimers so that they appear brighter compared to the spherical particles. This fact has been confirmed in the defocused images of the system where these bright particles appear with distinct interference fringe patterns. The scale length is 3 mm. (c) The side-view image of the quasi-2D suspension, which clearly shows the presence of dimers below the plane of spherical particles. Here, $\Delta_{md}$, $\Delta_{mm}$, and $\Delta_{dd}$ represent the separations between monomer(m)-dimer (d), monomer(m)-monomer(m), and dimer(d)-dimer(d), respectively. The deformation of the quasi-2D suspension due to the presence of dimers is measured by the angle $\theta$.}
\label{dimer_rotation}
\end{figure}

The experiments were performed with a (modified) Gaseous Electronics Conference (GEC) chamber, in a capacitively coupled rf glow discharge at 13.56 MHz (see Fig.~\ref{GEC}). The top and side windows of the chamber provide access for the video recording and laser illumination system. In the {\it injection} phase, the melamine formaldehyde particles with diameter of 7.1 $\mu$m have been inserted into the plasma using a dispenser at a neutral gas pressure of 0.66 Pa and rf power of 20 W. In this phase, along with spherical dust particles, the impurities in the form of heavy agglomerates or asymmetric particles are also present. They levitate below the monolayer of monodisperse spherical particles at different heights at the central part of the discharge chamber due to their charge-to-mass ratio. Using defocused imaging technique, it is found that there are mainly three types of particles in the system: monodisperse spherical particles, dimers with dark stationary/rotating interference fringes and much heavier particles without interference fringe patterns. Then neutral gas pressure is increased up to 1.34 Pa and ``purification'' process is started through which all these bigger/asymmetric particles can be dropped on the lower electrode. In case of ``complete'' purification, all these heavier particles should drop on the lower electrode and a monolayer of only one type of spherical particles can be formed (2D suspension). But, in case of ``partial'' purification, only heavier particles are removed and a quasi-2D suspension can be formed as a binary mixture of spherical particles and their binary agglomerates (dimers) as shown in Fig.~\ref{GEC}. In this case, the spherical particles form a quasi-monolayer above the central part of the rf electrode, with the striped particles levitating in a layer just below the monolayer with vertical separation, d $\ll$ $\Delta$, the interparticle separation in monolayer. However, the 2D suspension of spherical particles exists at the periphery. The particle suspension was illuminated with a horizontal sheet of red diode laser light (wavelength of 660 nm) and imaged through the top glass window with a Photron FASTCAM 1024 PCI camera operating at a speed of 60 frames/s with a field of view of 1024 x 1024 pixels. Both quasi-2D and 2D suspension have been imaged simultaneously. The camera lens was equipped with a narrow-band interference filter to collect only the illumination laser light scattered by the particles.

{\it Observation:}
Unlike spherical particles, the binary agglomerates (dimers) can be identified instantly by looking at their top-view defocused images which contain dark stationary/spinning interference fringe patterns (striped particles). This particular characteristic of the binary agglomerates has been used to create quasi-2D suspension in a controlled manner. After partial purification, the maximum intensity of the laser beam is placed at the plane of binary agglomerates. Then in the focused image, these particles appear brighter compared to the spherical particles. To check whether all these particles are indeed the binary agglomerates, we take defocused images after each purification. When all the brighter particles seem to contain rotating interference fringe patterns on their defocused images, we stop further purification. The defocused top-view images of such brighter particles are shown in Fig.~\ref{dimer_rotation}. Then we come up with quasi-2D suspension as a binary mixture of spherical particles and their binary agglomerates (dimers) at the central part of the electrode. It is found that the maximum number of dimers are present for 7 micron particles ($\approx 30-40 \%$) followed by 8 microns ($\approx 25 \%$), 11 microns ($\approx 15 \%$) and 9 micron particles ($< 5 \%$). It is found that the vertical separation between monomers and dimers can be estimated with an accuracy $\approx 10 \%$. Systematic investigations show that the dimers also influence the melting of crystal in the centre. It is very difficult to sustain quasi-2D suspension at lower pressure (below 1 Pa) when the number density of dimers is high ($ > 15 \%$) compared to monomers.  

\begin{figure}
\includegraphics[width=\linewidth]{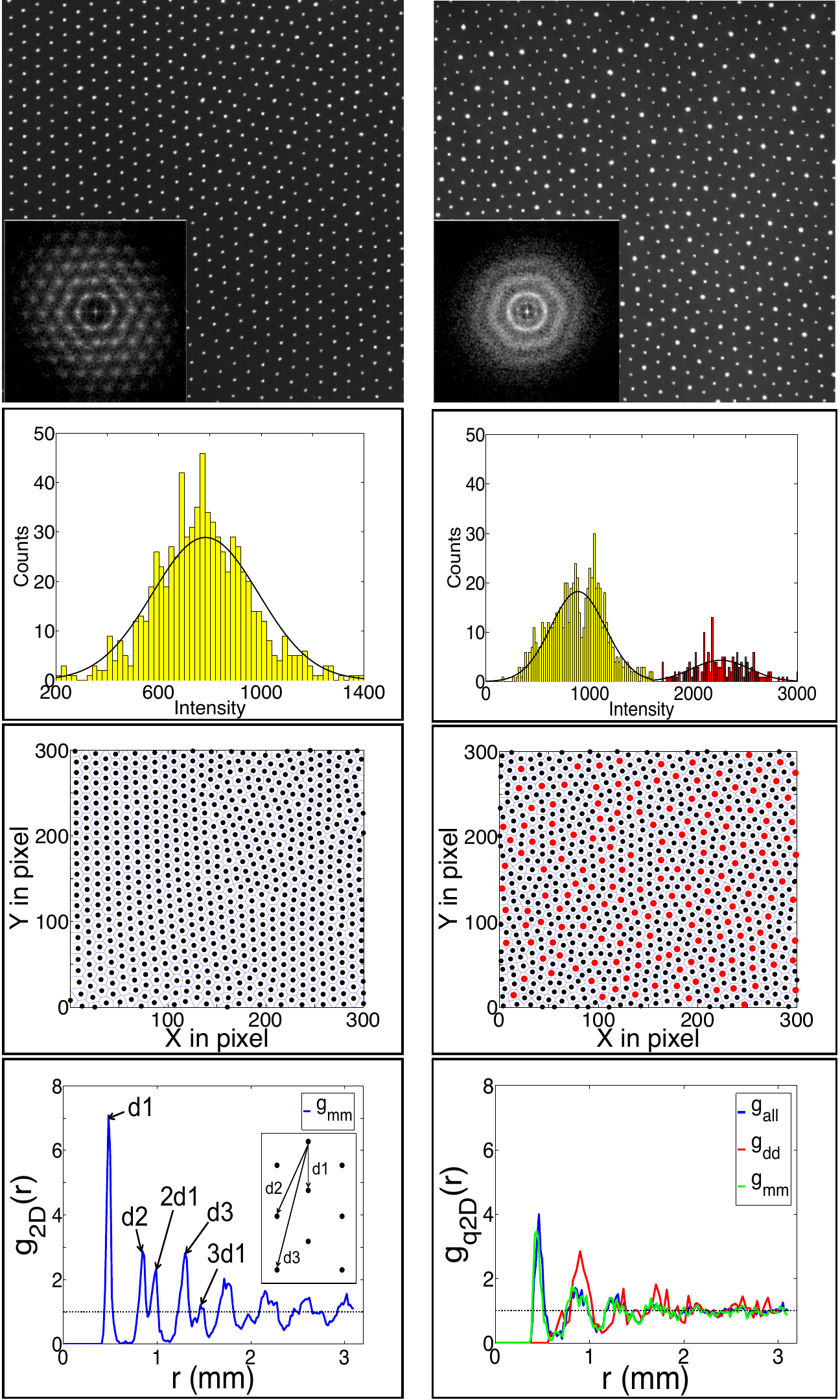}
\caption{Left panels represent the structural analysis of 2D suspension (ROI-1) and the right panels represent that of quasi-2D suspension as binary mixture of spheres and dimers (ROI-2). The experiment has been performed with 7 $\mu$m melamine formaldehyde (MF) particles in argon discharge at 1.34 Pa neutral gas pressure and 20 W rf power. The histogram plot of intensity profiles of ROI-1 has one peak indicating the presence of only one type of spherical particles (monomers), whereas in ROI-2 there are two peaks indicating the population of two different types of particles: monomers (yellow) and dimers (red). Gaussian fits have been performed in both cases. Based on this histogram profile, the dimers can be distinguished from the monomers and positions of two types of particles have been color-coded: monomers (black) and dimers (red). Voronoi diagrams have been computed along with the positions of the particles. The pair correlation function in ROI-1 indicates that the 2D crystal has almost perfect hexagonal symmetry with the splitting of second peak (inset). However, the binary mixture in ROI-2 represents disordered solid. In this case, $g(r)$ has been calculated for all three cases: monomer, dimer, and binary mixture of monomers and dimers. The presence of dimers reduces the magnitude of the first peak of g(r) to almost half of its value for ROI-1. The FFT of ROI-1 indicates that the system has an ordered crystalline structure, but in ROI-2 we have disordered solid. }
\label{structural_analysis} 
\end{figure}

The systematic analysis has been performed after partial purification as shown in Fig.~\ref{structural_analysis}. The first task to analyze quasi-2D suspension is to distinguish dimers from monomers. This has been achieved by taking top view image of the quasi-2D crystal where the laser is placed in the plane of dimers with maximum intensity so that dimers appear much brighter than the monomers. This is possible as the vertical separation between monomers and dimers is $d \sim 200  \mu$m which is larger than the laser beam width ($\sim 180  \mu m$). Then we plot the histogram of the intensity profile of such a top-view image which clearly shows two populations of particles with different intensities. Clearly, the less populated brighter particles are dimers and the highly populated but less bright particles are spheres. The two types of particles have been color coded: red particles with higher intensity are dimers and black particles with low intensity are spheres. The voronoi diagram has been made along with positions of particles. In 2D suspension, the voronoi cells are uniform, hexagonally symmetric. In quasi-2D suspension, a large number of defects (five-, seven-, eight-fold) appear along with hexagonal voronoi cells. However, in this case most of the hexagonal voronoi cells are deformed.  

\begin{figure}
\includegraphics[width=\linewidth]{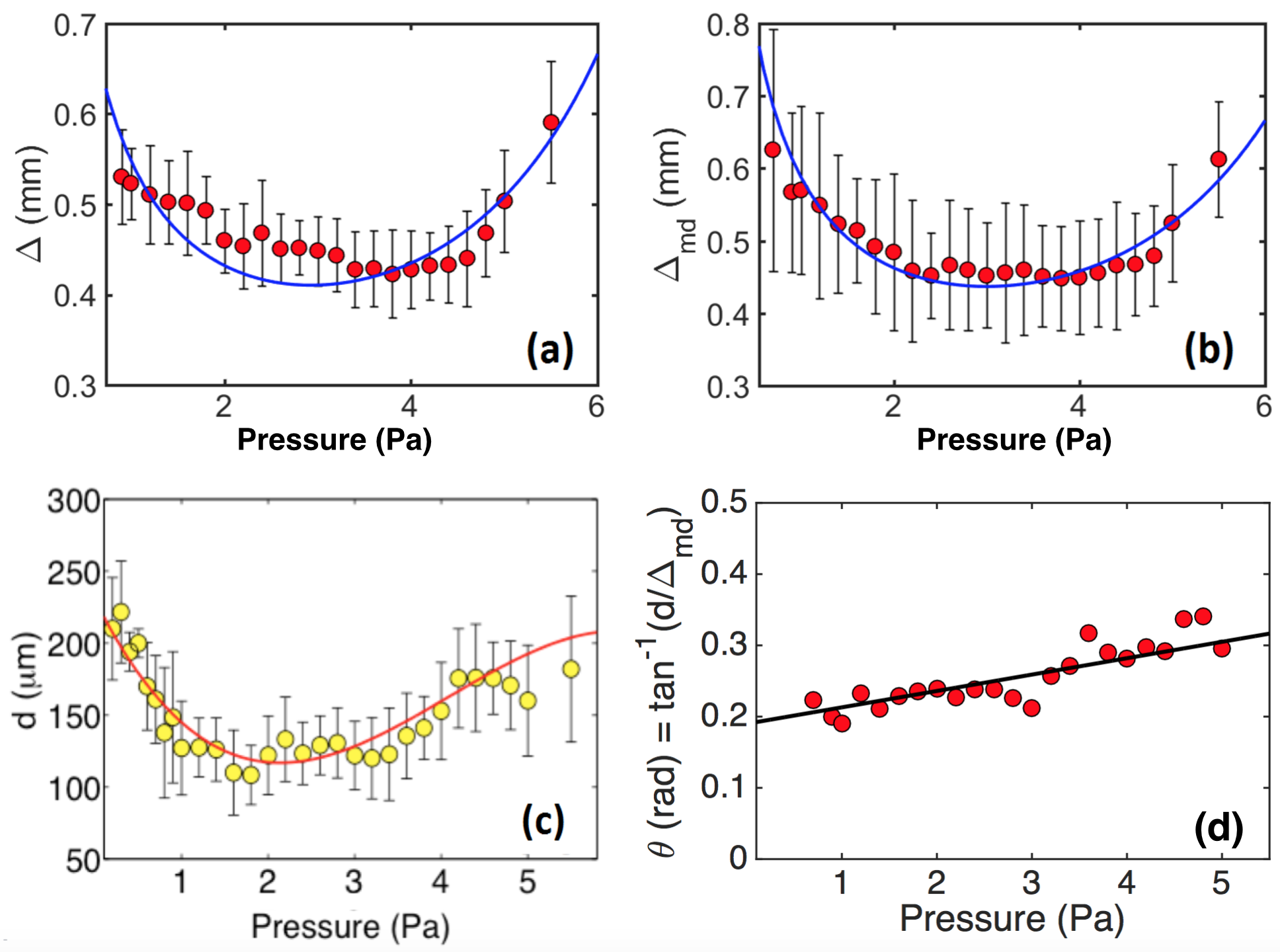}
\caption{(a) Pressure dependence of the interparticle distance 
$(\Delta)$ between particles (monomers) in ROI-1.~The symbols show the experimental data and the solid line shows the theoretical estimate given by Eq.~(\ref{interparticledistance}).~The theoretical curve was obtained using the following set of parameters: $T_{h}=300\,$K, $r_{p}$=3.5$\,\mu$m, $N_{p}$=1000, $a_{T}$=2.5, $b_{T}$=0.2 (see the description of the theoretical model).~(b) The pressure dependence of $\Delta_{\rm md}$, where 
$\Delta_{\rm md}$ is the projection of the distance between monomers and dimers on the horizontal plane of monomers.
The symbols show the experimental data and the solid line shows the theoretical estimate obtained using the following set of parameters: $T_{h}=300\,$K, $r_{p}$=3.5$\,\mu$m, $N_{p}$=1000, $a_{T}$=2.38, $b_{T}$=0.18.~Note also that $\Delta_{\rm md} \sim \Delta_{\rm m}$ at sufficiently high pressure ($p>3$~Pa). (c) The pressure dependence of the vertical separation ($d$) between the layers of monomers and dimers has been fitted with a cubic polynomial. (d) The pressure variation of the deformation of the crystal in ROI-2 can be scaled linearly as, $\theta = 0.023p + 0.19$.}
\label{deformation} 
\end{figure}

The local deformation in the disordered quasi-2D suspension exhibits the caged particle dynamics of dimers surrounded by five-, six-, seven-, eight-, etc. spherical particles with unequal bond lengths. The presence of dimers as caged particles increases with particle diameter. Sometimes the dimers also form hexagonal symmetry along with monomers. In the central part of the crystal, the dimers form islands surrounded by monomers. These are the local deformed structures created by the presence of dimers. The dimers may act as the local source of energy due to their dynamics which can be an alternate mechanism for crystal melting. The higher the number of dimers (more than 15 - 20 \%) in quasi-2D suspension, the easier it melts at low pressure (below 1 Pa).

To proceed with structural analysis for both systems, we computed radial distribution function $g(r)$ for a single image. The crystalline structure in 2D-crystal is evident from the first sharp peak of $g_{\rm 2D}(r)$ which corresponds to the position of nearest neighbors and the area of which gives the coordination number. In our case, the nearest interparticle distance d1 $\sim$ 481 $\mu$m. The characteristic splitting of second peak should correspond to the second (d2 = $\sqrt{3}{\rm d1} \sim {\rm 833} \mu$m) and third ($2{\rm d1}\sim {\rm 962} \mu$m) neighbor distances in the second coordination shell. In our case, the second and third peaks appear at $\sim$ 840 $\mu$m and $\sim$ 989 $\mu$m. The other peaks appear at d3 $\sim$ 1.298 $\mu$m and 3d1 $\sim$ 1.461 $\mu$m. At large distances, the peaks broaden as well as merge with each other and become lost in the continuum background $g(r) \rightarrow$ 1. The structural properties of $g(r)$ change drastically in quasi-2D suspension. The presence of dimers destroys the crystalline order made by the monomers. The height of the first peak of the $g_{q2D}(r)$ decreases by $\sim$ 40\%, although the position almost remains the same. The system approaches a fluid state. This can be attributed to the growing number of structural defects with increasing temperature. After distinguishing dimers from spheres, it is possible to compute $g(r)$ separately for monomers $g_m(r)$, for dimers $g_d(r)$, and for both, $g_{md}(r)$. The position of the first peak of $g_m(r)$ is $r_m \sim$ 440 $\mu$m, which is very close to that of $g_{\rm q2D}$(r) $\sim$ 463 $\mu$m. The first peak of $g_d(r)$, which is lower and wider also appears for dimers at $r_d \sim$ 900 $\mu$m. In this system, $g_{q2D}(r) \rightarrow$ 1 much faster than $g_{\rm 2D}(r)$, which implies that quasi-2D suspension (binary mixture of spheres and dimers) is much hotter than the 2D suspension made of monodisperse spheres. This is also evident from the FFT analysis of ROI's as shown in the Fig.~\ref{structural_analysis}.

The dependencies of in-plane interparticle separation $\Delta$ on neutral gas pressure (ion-neutral collisionality) have been measured in both ROI-1 and ROI-2 by looking at the position of the first peak of pair-correlation function, $g(r)$. The first peak of $g(r)$ has been fitted with gaussian distribution and the FWHM gives the errors in measurements. It is found that $\Delta$ initially decreases monotonically  with pressure in both cases but after reaching a minimum at $\sim 3.4$ Pa it increases again. However, it is to be noted that the two types of particles (monomers and dimers) are present in ROI-2 in different vertical planes, i.e., we can measure interparticle separation for spherical particles ($\Delta_{\rm m}$) and dimers ($\Delta_{\rm d}$) separately. It is to be noted that the local structure of monomers with at least one dimer is deformed in 3D structure. Since the vertical separation (d) $>$ laser width, we can illuminate both planes with maximum laser intensity. The vertical movement of the illumination laser can be monitored through a scale with minimum division of 10 $\mu$m. To achieve minimum error, each data point corresponds to $\sim$ 20 measurements. Then the average value and standard deviations have been calculated. But if we measure $\Delta_{\rm md}$ between spherical particles and dimers, then we should be careful as it represents the projection of actual $\Delta_{\rm md}$ on the plane of dimers. However, this measurement can be used to measure local deformation in ROI-2 if sphere-dimer vertical separation $d$ can be measured as well. The deformation in ROI-2 can be quantified in the form of angle $\theta = tan^{-1}(d/\Delta_{\rm md})$. The dependencies of $\Delta_{\rm md}$, $d$ and $\theta$ on neutral gas pressure are shown in Fig.~\ref{deformation}d. It is found that $d$ is independent of number of dimers. At low pressure (below 0.7 Pa), it can be measured in presence of very few dimers. In this regime, it is not possible to sustain quasi-2D crystal with significant dimers (more than 10 - 20 \%) as it melts. On the other hand, at higher pressure (above 3 Pa) the dimers start to disperse toward periphery. In this regime, $\Delta_{\rm md} \sim \Delta_{\rm m}$ and $d$ can be measured in presence of few dimers. In the intermediate regime ($\sim$ 0.7 - 3 Pa), dimer concentration is high in the central part of the electrode and we can measure $\Delta_{dd}$ using $g_d(r)$.\\ \indent
The experimental results can be interpreted using an estimate based on the theoretical model proposed by Totsuji et al.~\cite{totsuji2001structure}. Namely, let us consider a two-dimensional layer of dust particles which is formed
in a cylindrical discharge. Let us also make the following assumptions: the particles have a negative
charge with absolute value $Q$ and interact via a Yukawa potential with a screening length $\lambda$; the particles are distributed uniformly with a constant surface density; the transverse confinement potential can be written as 
$\phi(r) = -\omega r^2$, where $r$ is the distance from the discharge axis of symmetry. According to \cite{totsuji2001structure}, the
interparticle distance in this case is given by
\bee \label{interparticledistance}
\Delta=\lambda \left(\frac{4\alpha}{N_p}\right)^{1/4},
\ene
where $N_p$ is the number of particles and $\alpha$ is a dimensionless parameter given by $\alpha = (Q/\omega)\lambda^{-3}$. In typical experimental conditions, the screening length is close to the electron Debye length, i.e., $\lambda =\sqrt{4\pi n_e e^2/kT_e}$, where $n_e$ and $T_e$ are the electron density and temperature in plasma, respectively. The dust particle charge can be calculated as $Q = r_p |\phi_p|$, where $r_p$ is the particle radius and $\phi_p$ is the particle surface potential obtained from the orbital motion limited theory. \\ \indent
The confinement stiffness $\omega$ can be estimated using the following considerations. The distribution of the electric potential $\phi$ near the axis of symmetry can be found from the Boltzmann relation
\bee \label{pot}
\phi = -(kT_e/e) \ln(n_e/n_{e0}),
\ene
where $n_{e0}$ is the electron concentration on the axis of symmetry. The distribution of $n_{e}$ near the axis of symmetry can be found from the solution of the drift-diffusion equations. This well-known solution is given by
\bee \label{elecdensity_axis}
n_e = n_{e0}\,J_0(r/\ell_0),
\ene
where $J_0$ is the zero order Bessel function and $\ell_0$ is the characteristic length defined as
\bee \label{mfp}
\ell_0 = \frac{v_{b}}{n_{a}\sqrt{K_{i}\nu_{ia}}},
\ene
where $v_b$ is the Bohm velocity, $n_a$ is the concentration of neutrals, $K_i$ and $\nu_{ia}$ are the ionization rate and ion-atom momentum exchange rate, respectively.~Here, the Bohm velocity is defined as $v_{b}=\sqrt{(kT_{e}+kT_{h})/m_{h}}$, where $T_{h}$ and $m_{h}$ are the temperature and mass of heavy particles (ions and atoms), respectively. The ionization rate $K_{i}$ can be evaluated using the corresponding expression for the direct ionization process presented, e.g., in \cite{almeida2000simulation}. The ion-atom momentum exchange rate is given by 
$\nu_{ia}=(8/3)\sqrt{kT_{h}/\pi m_{h}}\bar{Q}_{ia}^{(1,1)}$, where $\bar{Q}_{ia}^{(1,1)}$ is the averaged momentum-transfer cross section. For argon, this cross-section can be evaluated using the results of \cite{devoto1973transport}.
\\ \indent
For typical conditions, the radial size of the layer is several times smaller than $\ell_0$. In this case the Bessel function $J_0$ in Eq.~(\ref{elecdensity_axis}) can be approximated as
\bee \label{besselfunction}
J_0(r/\ell_0) \approx 1 - 0.25(r/\ell_0)^2.
\ene
Using Eqs.~(\ref{pot}),~(\ref{elecdensity_axis}) and (\ref{besselfunction}), one can obtain the following expression for the potential distribution near the axis:
\bee \label{potdist}
\phi \approx -(1/4)\,T_e \,(r/\ell_0)^2.
\ene
Therefore, the stiffness parameter $\omega$ is given by
\bee \label{freq}
\omega = (1/4)\,T_e\,\ell_{0}^{-2}.
\ene
Different diagnostics show that the pressure dependence of electron temperature and density can be scaled linearly: $n_e = a_n + pb_n$ and $T_e = a_T - pb_T$ where $n_e$ is in $10^8\,\mathrm{cm}^{-3}$ and $p$ is in Pa. The values of the coefficients $a_n, b_n, a_T$ and $b_T$ varies over discharge conditions. For DC discharge plasmas in the weakly collisional regime, $a_n = 0.9, b_n = 0.03, a_T = 8.3, b_T = 0.02$~\cite{Khrapak_PhysRevE.72.016406} and for highly collisional regime, $a_n = 4.45, b_n = 0.034, a_T = 7.54, b_T = 0.0067$~\cite{Khrapak_EPL_2012}. For RF discharge plasmas, $a_n = 20$ and $b_n = 12$~\cite{Lenaic}.
However, it is found that the pressure variations of $\Delta$ and $\Delta_{md}$ in our work are not sensitive to electron density but depends strongly on $T_e$.~No experimental data is available for $T_e$, but the coefficients $a_T$, $b_T$ and $N_p$ can be adjusted numerically in such a way that the pressure dependence of the interparticle distance is well described by Eq.~(\ref{interparticledistance}). As an example, we show in Fig.~\ref{deformation}a the comparison between our theoretical estimates and the experimental data obtained for $\Delta$ in ROI-1. It can be seen in Fig.~5(a) that our theoretical model is in reasonable agreement with the experimental measurements. It is interesting to note that the pressure dependence of the projection 
$\Delta_{\mathrm{md}}$ in ROI-2 is also described reasonably well by Eq.~(\ref{interparticledistance}) (see Fig.~\ref{deformation}b), despite the fact that this equation was originally derived for monomer systems.
\\ \indent
In conclusion, the structural analysis of quasi-2D suspension (binary mixture of spheres and dimers) has been presented and compared with 2D suspension made of spherical particles. First, we have used systematic partial purification to create quasi-2D suspension in the centre and 2D suspension at the periphery of the same system of investigation. A defocused imaging technique is used to identify binary agglomerates (dimers) which have a rotating interference fringe pattern on that defocused image, unlike spherical particles. Then, taking advantage of higher vertical separation between monomers and dimers compared to laser sheet thickness, we have identified and separated the dimers at the individual particle level. It is found that the presence of dimers destroys the crystalline order made by the spherical particles by creating defects and the system enters into a disordered solid state as evident from the radial distribution function and structure factor analysis. We have observed nonmonotonic dependences of in-plane interparticle separation, sphere-dimer separations, as well as vertical sphere-dimer separations. The experimental measurements of in-plane interparticle separations agree well with theoretical predictions. However, to explain the dependence of vertical separation on pressure, the wake effect should be taken into account, which is beyond the scope of present paper and left for future work.  \\

{\bf Acknowledgement:} M.C is supported by Marie-Curie international outgoing fellowship (IOF) under EU 7th framework programme.

\end{document}